\documentclass{article}

\usepackage{PRIMEarxiv}

\usepackage[utf8]{inputenc} 
\usepackage[T1]{fontenc}    
\usepackage{hyperref}       
\usepackage{url}            
\usepackage{booktabs}       
\usepackage{amsfonts}       
\usepackage{nicefrac}       
\usepackage{microtype}      
\usepackage{lipsum}
\usepackage{fancyhdr}       
\usepackage{graphicx}       
\usepackage{cite}
\usepackage{amsmath,amssymb,amsfonts}
\usepackage{graphicx}
\usepackage{textcomp}
\usepackage{multirow}
\usepackage{threeparttable}
\usepackage{booktabs}
\usepackage{subfigure}
\usepackage{algorithm,algpseudocode}
\usepackage[algo2e]{algorithm2e} 
\usepackage{adjustbox}
\graphicspath{{media/}}     

\title{Semi-Supervised Semantic Segmentation of Cell Nuclei via Diffusion-based Large-Scale Pre-Training and Collaborative Learning
}

\author{
  Zhuchen Shao \\
  Electronic and Information Engineering \\
  Tsinghua university, China \\
   \And
   Sourya Sengupta \\
  Department of Elec. \& Computer Eng. \\
  University of Illinois Urbana-Champaign, IL, USA \\
   \And
   Hua Li \\
   Department of Bioengineering\\
   University of Illinois Urbana-Champaign, IL, USA\\
  Department of Radiation Oncology  \\
  Washington University in St. Louis \\
   \And
   Mark A. Anastasio \\
  Department of Bioengineering \\
  Department of Elec. \& Computer Eng.\\
  University of Illinois Urbana-Champaign, IL, USA \\
  \texttt{maa@illinois.edu} \\
}

\begin{document}
\maketitle

\begin{abstract}
Automated semantic segmentation of cell nuclei in microscopic images is crucial for disease diagnosis and tissue microenvironment analysis. Nonetheless, this task presents challenges due to the complexity and heterogeneity of cells. While supervised deep learning methods are promising, they necessitate large annotated datasets that are time-consuming and error-prone to acquire. Semi-supervised approaches could provide feasible alternatives to this issue. However, the limited annotated data may lead to subpar performance of semi-supervised methods, regardless of the abundance of unlabeled data. In this paper, we introduce a novel unsupervised pre-training-based semi-supervised framework for cell-nuclei segmentation.
Our framework is comprised of three main components. Firstly, we pretrain a diffusion model on a large-scale unlabeled dataset. The diffusion model's explicit modeling capability facilitates the learning of semantic feature representation from the unlabeled data. Secondly, we achieve semantic feature aggregation using a transformer-based decoder, where the pretrained diffusion model acts as the feature extractor, enabling us to fully utilize the small amount of labeled data. Finally, we implement a collaborative learning framework between the diffusion-based segmentation model and a supervised segmentation model to further enhance segmentation performance.
Experiments were conducted on four publicly available datasets to demonstrate significant improvements compared to competitive semi-supervised segmentation methods and supervised baselines. A series of out-of-distribution tests further confirmed the generality of our framework. Furthermore, thorough ablation experiments and visual analysis confirmed the superiority of our proposed method.

	\end{abstract}
	
	
	\section{Introduction}
Precise segmentation of cell nuclei reveals important cellular features and helps with cancer grading and prognostic prediction and analyzing cell type interactions \cite{chen2020pathomic, ahmedt2022survey}. 
However, cell segmentation from microscopy images can be challenging due to complex cellular structure and close proximity or overlap between cells.  
Recently, supervised deep learning methods have emerged as crucial tools for cell nuclei segmentation \cite{janowczyk2016deep, srinidhi2021deep}. Supervised methods require a significant amount of annotated data to train deep learning models. However, due to the high-resolution and wide-field-of-view characteristics of microscopic images, manual annotation of gigapixel images that contain a large number of nuclei is time-consuming and error-prone \cite{graham2019hover}. 

Semi-supervised semantic segmentation utilizes a relatively small number of labeled data and a larger amount of unlabeled data for segmentation \cite{pelaez2023survey,jiao2022learning}. Various such semi-supervised methods adopt adversarial learning approaches \cite{zhang2020robust, zhang2017deep}, consistency regularization \cite{ouali2020semi, sengupta2023semi,yu2019uncertainty}, or pseudo-labeling~\cite{yang2022st++, chen2021semi,    qiao2018deep}. However, the limited availability of annotated data can negatively impact the performance of the semi-supervised methods, even when a large amount of unlabeled data is available. 
One way of mitigating this is to learn efficient data embedding by using large-scale unsupervised pre-training using generative models \cite{shwartz2022we,devlin2018bert, he2022masked, zhou2022self}. 
Diffusion models have emerged as the state-of-the-art technique for generative modeling \cite{ho2020denoising, dhariwal2021diffusion,nichol2021improved}. Additionally, diffusion models have been demonstrated to learn meaningful semantic information~\cite{baranchuk2021label,devlin2018bert,chen2020big}.
As a result, unsupervised pre-training of diffusion models holds promise for enhancing the performance of semi-supervised semantic segmentation.

While the previous works achieved promising results, they did not involve large-scale unsupervised pre-training for biomedical imaging applications and therefore did not address certain important domain-specific issues. Firstly, the effectiveness of semi-supervised segmentation methods that utilize a diffusion model-based large-scale pre-training to learn semantic feature embeddings when microscopy images are considered remains to be assessed. Secondly, in the  applications addressed previously, it was assumed that a large ensemble of images was  available for unsupervised pre-training and that this ensemble was representative of the images to-be-segmented at inference time. However, for biomedical applications such as cell nuclei segmentation, 
a sufficiently large ensemble of unlabeled images  may not always be available. Therefore, there remains an important need to systematically evaluate semi-supervised methods for biomedical applications that: 1) utilize semantic feature embeddings established by large-scale unsupervised pre-training of diffusion models that are trained by the use of limited training data that are representative of the to-be-segmented images, and 2) utilize semantic feature embeddings established by large-scale unsupervised pre-training of diffusion models that are trained by using large ensembles of unlabeled data that are not representative of the images to be segmented (out-of-distribution case, OOD).


To address the challenges possessed by traditional semi-supervised methods, this work investigates a large-scale pre-training-based novel semi-supervised framework for cell nuclei segmentation. The proposed framework comprises the following steps: \textbf{1)}~pre-train a diffusion model with an unlabeled set of images, 
\textbf{2)}~extract semantic features using the pre-trained diffusion model, \textbf{3)}~exploit these semantic features to predict segmentation labels using a transformer-based decoder and a segmentation head. To address the issues of limited pre-training data and out-of-distribution cases, we also incorporated collaborative learning \cite{song2018collaborative, guo2020online, zhou2019collaborative}, combining traditional semantic segmentation approaches with the proposed diffusion-based framework.
We performed comprehensive numerical experiments on four publicly available datasets for cell nucleus semantic segmentation. The results demonstrate that our proposed model leads to significant improvements compared to other semi-supervised methods and supervised baselines. 

The main contributions of our work include: 

\textbf{1)}  To the best of our knowledge, this is the first work to demonstrate how the diffusion model’s semantic feature learning capabilities can be exploited with large-scale unsupervised pre-training for semi-supervised cell nuclei segmentation.  We show diffusion models are strong semi-supervised learners even when the downstream to-be-segmented dataset is not available during the large-scale pre-training. 

\textbf{2)} We show that collaborative learning can further improve the segmentation performance of the proposed framework when pre-training data are limited, and when the to-be-segmented dataset is OOD (not available during pre-training). As a ‘good collaborator’, the diffusion pre-training-based framework can be effectively combined with the supervised segmentation frameworks to enhance performance. 	

{The remainder of the paper is organized as follows: Section II provides the background information, and in Section III, the proposed method is described. Section IV describes the experimental setting, including the dataset used, evaluation metrics, and implementation details. The experimental results and their analysis are presented in Section V through quantitative and qualitative assessments. Finally, Section VI concludes the paper.}

	\section{Background}

\subsection{Semi-supervised Cell Nuclei Segmentation}
Supervised deep learning models for cell nuclei segmentation demands a large amount of pixel-level annotation by experts which can be labor-intensive and error-prone. To alleviate this challenge, semi-supervised segmentation algorithms have emerged as a promising approach, leveraging a limited annotated data along with a larger amount of unlabeled data ~\cite{jin2023label, hollandi2022nucleus,sahasrabudhe2020self}.
Current approaches include methods such as consistency learning and pseudo-labeling~\cite{jiao2022learning}. Consistency regularization approaches encourage model predictions to be consistent under different perturbations ~\cite{sengupta2023semi, wu2022cross}, promoting robustness. Pseudo-labeling approaches assign pseudo-labels to unlabeled samples based on model predictions, enabling their inclusion in the training process iteratively~\cite{li2023semi}. However, limited annotated data may result in poor performance of semi-supervised methods, regardless of the availability of a large amount of unlabeled data. 
Unsupervised large-scale pre-training with unlabeled data can be an alternative framework in semi-supervised semantic segmentation.
 
\subsection{Diffusion Model}	
Diffusion models are state-of-the-art generative models that have been widely used in various fields and outperformed other generative models in terms of the generated high-quality images \cite{kazerouni2023diffusion, croitoru2023diffusion, shao2023augdiff}. The denoising diffusion probabilistic model (DDPM) \cite{ho2020denoising} is a well-established diffusion model. Here the diffusion process is accomplished through two fundamental stages:  \textbf{1)} forward diffusion process, where noise is gradually introduced to the input data, and the noise level is systematically increased until the data is transformed into pure Gaussian noise and \textbf{2)} reverse diffusion process, where the original structure of the data is restored from the perturbed distribution using a denoising process.   Recently, it has been discovered that DDPM excels not only in generating high-quality images but also in learning valuable semantic feature embeddings from training data.\cite{baranchuk2021label}. 

However, implementing DDPM poses challenges due to its lengthy sampling times, high computational costs, and the significant training data it requires. To address these limitations, latent diffusion models have recently been introduced~\cite{nichol2021improved}. These models employ a compression method to generate latent representations of images for the training of the diffusion model.  This involves leveraging a pre-trained autoencoder, where the encoder generates the latent representation prior to the forward diffusion, and the decoder reconstructs the final image following the backward diffusion.  The backward diffusion process uses a UNet model with a cross-attention mechanism. Specifically, the compression technique reduces the computational burden and minimizes the need for extensive training data compared to the original DDPM model. The cross-attentive UNet model significantly enhances the quality and distribution of generated images. The stride sampling steps proposed in \cite{song2020denoising} also greatly reduce the lengthy sampling times of latent diffusion compared to DDPM. Moreover, latent diffusion has demonstrated effective feature learning capability through unsupervised training in various tasks \cite{shao2023augdiff, zeng2022lion}.

	\begin{figure*}[!t]
		\centerline{\includegraphics[width=\linewidth]{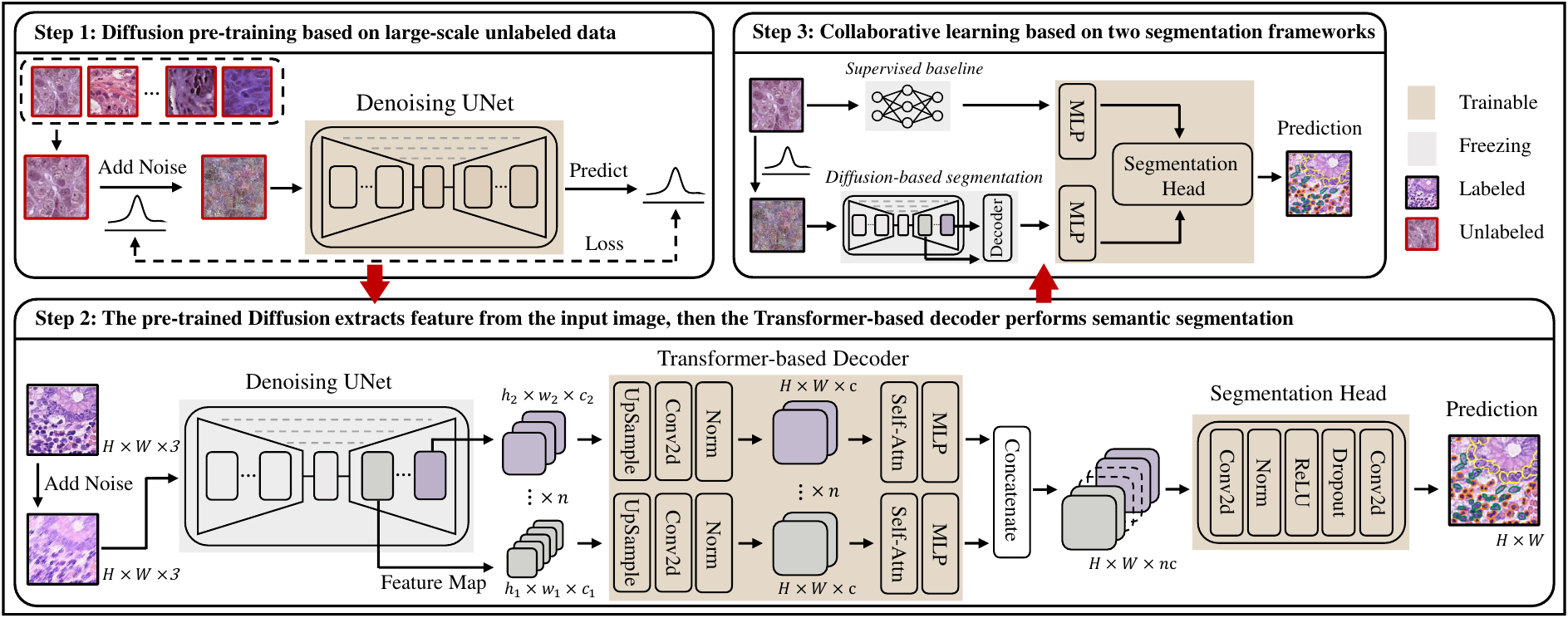}}
		\caption{\textbf{Overview of our proposed framework.} 
			\textbf{Step 1.} Pre-training of the diffusion model using large-scale unlabeled data, with a generation task aiding in learning semantic information from the cell nuclei images.
			\textbf{Step 2}: Using the pre-trained diffusion model to extract semantic features of cell nuclei images, feature maps from different blocks of denoising unet are aggregated using a transformer-based decoder that predicts the cell nucleus semantic segmentation results.
			\textbf{Step 3}: The diffusion-based segmentation framework is trained and combined with traditional semantic segmentation to reduce the generalization error when tested on out-of-distribution data or when diffusion is pre-trained with limited unlabeled data.
		}
		\label{figure_overview}
	\end{figure*}

\subsection{Collaborative Learning}
Collaborative learning is a framework that aims to enhance overall performance and mitigate potential performance drop or generalization errors by leveraging the collaboration of multiple models~\cite{song2018collaborative,guo2020online,zhou2019collaborative}.
Song \textit{et~al.}~\cite{song2018collaborative} introduced this method in deep learning to improve efficiency against label noise in image classification tasks. 
Guo \textit{et~al.}~\cite{guo2020online} employed collaborative learning to develop an online distillation approach in knowledge distillation, which effectively enhanced the performance of the unilateral distillation method when dealing with input domain perturbations.
Zhou \textit{et~al.}~\cite{zhou2019collaborative} utilized collaborative learning to leverage different types of annotated data for multi-task predictions.

	\section{Proposed Method} 
The proposed framework consists of the following steps. Primarily a diffusion model was pre-trained with large-scale unlabeled cell nuclei images.  Secondly, semantic features were aggregated from the pre-trained diffusion model using a transformer-based decoder to get the segmented outputs. To further improve the performance of the framework under limited pre-training data and out-of-distribution (OOD) cases, collaborative learning was introduced to reduce potential performance drops. The proposed framework is henceforth named as diffusion encoder-transformer decoder-based segmentation framework (DTSeg).   A comprehensive overview of the proposed framework can be seen in Figure~\ref{figure_overview}.

	
	\subsection{Diffusion-based Large-scale Pre-training}

 The primary step of the proposed DTSeg method employed large-scale unsupervised pre-training to learn efficient feature embedding from the cell nuclei images.  The latent diffusion model \cite{Rombach_2022_CVPR} was chosen for large-scale pre-training due to the advantages of this method, described in Section II.B.  As described in Section II.B, during the training process, $x\in\mathbb{R}^{H\times W\times3}$, the encoder $\mathcal{E}$ of the latent diffusion model encoded a given a cell nuclei image into a latent representation $z = \mathcal{E}(x)$, where $z\in\mathbb{R}^{\frac{H}{f}\times \frac{W}{f}\times3}$ where $f$ denotes the downsampling factor, $H$ and $W$ denote image height and width respectively. The decoder $\mathcal{D}$ reconstructed the cell nuclei image from the latent space, resulting in $\tilde{x}=\mathcal{D}(z)=\mathcal{D}(\mathcal{E}(x))$, where $\tilde{x} \in \mathbb{R}^{H\times W\times3}$. Therefore, the pre-trained image compression networks $\mathcal{E}$ and $\mathcal{D}$ facilitated diffusion training in a more efficient and lower-dimensional latent space. Note that we employed the large-scale pre-trained autoencoder provided in \cite{Rombach_2022_CVPR} to obtain $\mathcal{E}$ and $\mathcal{D}$.
In the forward diffusion process, Gaussian noise $\boldsymbol{\epsilon}$ was continually added to the input $z$ in total $T$ steps to create a sequence of noisy samples $\{z_{t}\}_{t=1}^{T}$. In contrast, the reverse process involved the denoising Unet ($\boldsymbol{\epsilon}_\theta$) for predicting the noise from the noisy cell nuclei image. 
The schematic of the step is shown in step 1 of Fig.~\ref{figure_overview}.
	

	\subsection{Diffusion-based Semantic Segmentation Framework}
 The purpose of this step was to obtain feature maps from different blocks of the denoising UNet of the pre-trained diffusion in order to capture semantic information from various layers.  To maximize the utilization of semantic features extracted by the pre-trained diffusion $\boldsymbol{\epsilon}_\theta$, a transformer-based decoder was proposed that can simultaneously aggregate semantic features from different blocks of the denoising UNet.
Transformer is a highly effective feature aggregation technique that has applications in various deep learning tasks \cite{dosovitskiy2020image, shao2021transmil}. By leveraging its powerful self-attention mechanism, the network was able to aggregate intrinsic features of intermediate layers of UNet.  The feature maps $\{f_i\}_{i=1}^{n}$ obtained from the intermediate layers of Unet had varying sizes. Therefore, upsampling layers followed by convolution layers were employed to ensure the same size of all the feature maps as the semantic segmentation label $y$. 

The output of the transformer-based decoder was passed through a shallow 2-layer segmentation head for the final segmentation prediction.  
The schematic of the step is shown in step 1 of Fig.~\ref{figure_overview}.

The detailed training process for the  proposed approach, DTSeg, is presented in Algorithm~\ref{alg:train}.

	\begin{algorithm}[ht]
		\small
		\caption{Training Process of DTSeg}
		\label{alg:train}
		\KwIn{
			Input image $x$, semantic segmentation label $y$, where $x\in\mathbb{R}^{H\times W\times3}$, $y\in\mathbb{R}^{H\times W}$. Pre-trained diffusion model $\boldsymbol{\epsilon}_\theta$. 
			
		}
		\KwOut{Trained transformer-based decoder.}
		\While{not converged}{
			
			{\%\textbf{1.} Encoder $\mathcal{E}$ encodes the image $x$ into latent representation.}\\
			$z = \mathcal{E}(x)$ \\
			{\%\textbf{2.} Randomly sample noise from Gaussian distribution.}\\
			$\boldsymbol{\epsilon} \sim \mathcal{N}(\mathbf{0}, \mathbf{I})$ \\
			{\%\textbf{3.} Take a small step $t$ and apply the diffusion process.}\\
			${z}_t \leftarrow \prod \limits_{i=1}^t q\left({z}_i|{z}_{i-1}\right)$ \\
			{\%\textbf{4.} Pre-trained diffusion $\boldsymbol{\epsilon}_\theta$ is used for noise prediction, and its feature maps $\{f_i\}_{i=1}^{n}$ of intermediate layers are extracted for downstream segmentation tasks.}\\
			$\{f_i\}_{i=1}^{n} \leftarrow \boldsymbol{\epsilon}_\theta\left({z}_t, t\right)$, $f_i\in\mathbb{R}^{h_i\times w_i\times c_i}$ \\
			{\%\textbf{5.} Preprocessing of feature maps $\{f_i\}_{i=1}^{n}$ to ensure all feature maps have the same size.}\\
			$\{F_i\}_{i=1}^{n} \leftarrow \operatorname{Conv2d}\left(\operatorname{Upsample}(\{f_i\}_{i=1}^{n})\right)$, $F_i\in\mathbb{R}^{H\times W\times c}$  \\
			{\%\textbf{6.} Using a transformer layer to learn the semantic information of feature maps $\{F_i\}_{i=1}^{n}$.}\\
			$\{F_i^T\}_{i=1}^{n} \leftarrow \operatorname{MLP}\left(\operatorname{SelfAttn}\left(\{F_i\}_{i=1}^{n}\right)\right)$, $F_i^T\in\mathbb{R}^{H\times W\times c}$  \\
			{\%\textbf{7.} Concatenating different feature maps along the channel dimension and obtaining the final prediction through the segmentation head.}\\
			$\hat{y} \leftarrow \operatorname{Head}(\operatorname{Concat}(\{F_i^T\}_{i=1}^{n}))$  \\
			{\%\textbf{8.} Update the transformer-based decoder using gradient descent with the dice loss function as the optimization objective.}\\
			$		\nabla \left\{\operatorname{Dice}(\hat{y}, y)\right\}
			$
		}
	\end{algorithm}

	\subsection{Step 3: Collaborative Learning Framework}
The purpose of this step was to improve the proposed DTSeg's performance in domain-specific situations, such as limited unlabeled pre-training datasets and out-of-distribution (OOD) cases. To address these challenges, a collaborative learning-based training strategy was proposed.  Collaborative learning involves jointly training models by leveraging the strengths of different participants, offering a promising approach to effectively improve the overall framework's performance in specific scenarios.

As depicted in Step 3 of Fig.~\ref{figure_overview}, in the proposed approach, the features from the pre-trained diffusion-based model (gray block) were combined with the features extracted by a trained supervised segmentation model (gray block) to train a new segmentation head (brown block). 
Throughout the collaborative learning process, both the trained supervised model and the pre-trained diffusion-based models were kept fixed and used exclusively for feature extraction purposes. 
Subsequently, these extracted features were passed through the segmentation head to make the final predictions.
Specifically, the supervised baseline utilized in our study was trained with a limited amount of labeled data. the supervised model employed ResNet34 \cite{he2016deep} as the encoder and FPN \cite{lin2017feature} as the decoder. Additionally, the MLPs used for feature mapping in this step were single-layer neural networks.
More detailed training procedure for collaborative learning is demonstrated in Algorithm \ref{alg:collaboration}.

	\begin{algorithm}[ht]
		\small
		\caption{Training Process of Collaborative Learning}
		\label{alg:collaboration}
		\KwIn{
			Input image $x$, semantic segmentation label $y$, where $x\in\mathbb{R}^{H\times W\times3}$, $y\in\mathbb{R}^{H\times W}$. Pre-trained DTSeg $\mathcal{D_T}$. Pre-trained supervised baseline $\mathcal{S}$. 
		}
		\KwOut{Trained feature mapping network and segmentation head.}
		\While{not converged}{
			
			{\%\textbf{1.} Using two pre-trained models for feature extraction.}\\
			$f_D = \mathcal{D_T}(x)$, $f_D\in\mathbb{R}^{H\times W\times C_D}$ \\
			$f_S = \mathcal{S}(x)$, $f_S\in\mathbb{R}^{H\times W\times C_S}$  \\
			{\%\textbf{2.} Using different MLPs for feature mapping.}\\
			$F_D = \operatorname{MLP}\left(f_D\right)$, $F_D\in\mathbb{R}^{H\times W\times C}$ \\
			$F_S = \operatorname{MLP}\left(f_S\right)$, $F_S\in\mathbb{R}^{H\times W\times C}$ \\
			{\%\textbf{3.} Concatenating $F_D$ and $F_S$ along the channel dimension and obtaining the final prediction through the segmentation head.}\\
			$\hat{y} \leftarrow \operatorname{Head}(\operatorname{Concat}(F_D, F_S))$  \\
			{\%\textbf{4.} We optimize the $\operatorname{MLPs}$ and $\operatorname{Head}$ using gradient descent, with the objective being to minimize the Dice loss function.}\\
			$		\nabla \left\{\operatorname{Dice}(\hat{y}, y)\right\}
			$
		}
	\end{algorithm}
	
	\section{Numerical Studies}
	\subsection{Datasets}

\begin{figure}[h]
		\centerline{\includegraphics[width=\linewidth]{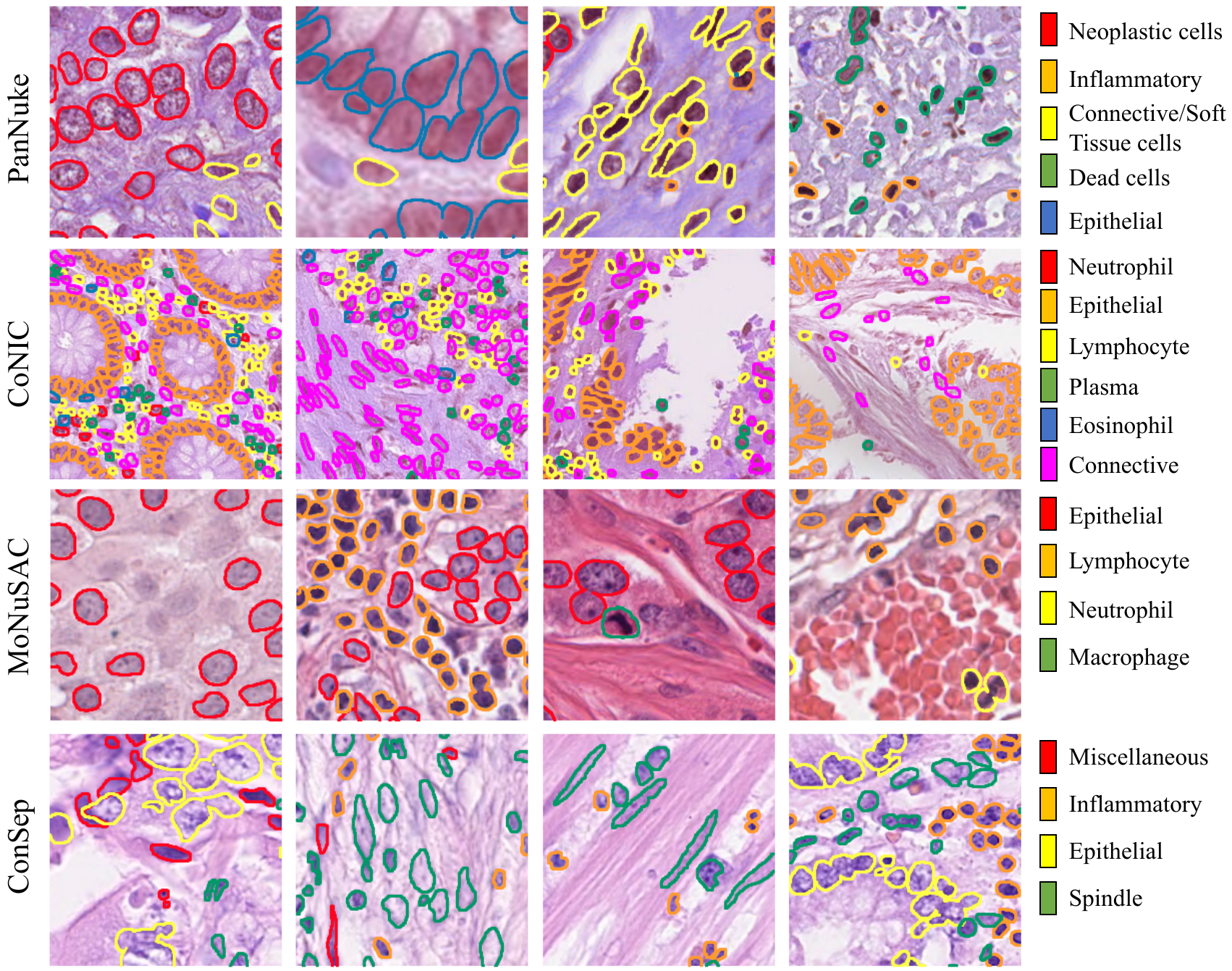}}
		\caption{
			\textbf{Visualization of the dataset.} Example images from PanNuke, CoNIC, MoNuSAC and ConSep are provided, along with their corresponding labels for cell nucleus semantic segmentation.
		}
		\label{figure_visualization_dataset}
	\end{figure}

For nuclei semantic segmentation, four publicly available datasets were used, namely PanNuke \cite{gamper2020pannuke}, 
 CoNIC \cite{graham2021conic}, 
 MoNuSAC \cite{verma2021monusac2020}, 
and ConSep \cite{graham2019hover}. 
The dataset details can be found in Table~\ref{tab:dataset}, and the annotated images with their corresponding semantic class labels are shown in Fig.~\ref{figure_visualization_dataset}. During the preprocessing of ConSep and MoNuSAC datasets, the images were cropped and resized. To ensure the diffusion model's generalization capability for out-of-distribution (OOD) cases, images from ConSep and PanNuke datasets were excluded from the CoNIC dataset, which originally contained images from various datasets. Table~\ref{tab:pretrained} provides information about the three diffusion pre-training dataset settings. Specifically, ``DTSeg (MoNuSAC)" and ``DTSeg (PanNuke)" indicate diffusion pre-training conducted solely with MoNuSAC or PanNuke datasets, respectively. For ``DTSeg (Big)", all three datasets were combined for the diffusion pre-training. Similarly, models denoted as ``Collaboration (MoNuSAC)", ``Collaboration (PanNuke)", and ``Collaboration (Big)" indicate collaborative learning models utilizing DTSeg (MoNuSAC), DTSeg (PanNuke), and DTSeg (Big), respectively.
	
	\begin{table}[h]
		\centering
		\setlength{\tabcolsep}{0.8mm}
		\caption{
			Summary of datasets used in study, including original and preprocessed information.
		}
		\label{tab:dataset}
		\begin{tabular}{c|c|c|c|c}
			\toprule
			& PanNuke \cite{gamper2020pannuke} & CoNIC  \cite{graham2021conic} & MoNuSAC \cite{verma2021monusac2020}            & ConSep \cite{graham2019hover}    \\ \midrule
			\#image                 & 7,901 & 4,805   & 584                  & 41        \\
			\#nuclei            & 189,744  & 549,108  & 46,909              & 24,319      \\
			magnification       & 20$\times$ or 40$\times$ & 20$\times$  & 40$\times$                 & 40$\times$   \\
			\#nuclear types       & 5  & 6      & 4                  & 4         \\
			image size   & 256$\times$256 & 256$\times$256 & \begin{tabular}[c]{@{}c@{}}90$\times$98 to \\  1,422$\times$2,162\end{tabular}  & 1,000$\times$1,000    \\
			\#organs           & 19 & 1         & 4                     & 1       \\  \midrule \midrule
			\#patch             & 7,901& 4,805     & 2,597                & 1,025        \\
			patch size  & 256$\times$256   & 256$\times$256    & 256$\times$256        & 256$\times$256      \\ \bottomrule
		\end{tabular}
	\end{table}
	
	\begin{table}[]
		\centering
		\caption{
			Summary of all pre-trained diffusion models, including the size of the pre-training dataset and its source(s).
		}
		\label{tab:pretrained}
		\begin{tabular}{c|c|c}
			\toprule
			& \#patch & Pre-training dataset                                      \\ \midrule
			Diffusion (MoNuSAC) & 1,750    & Training set of MoNuSAC                                                                                              \\ \midrule
			Diffusion (PanNuke) & 2,523    & Fold 2 of PanNuke                                                                                                    \\ \midrule
			Diffusion (Big)     & 10,326   & \begin{tabular}[c]{@{}c@{}}Training set of ConSep +\\ Training set of MoNuSAC +\\ Fold 1,2,3 of PanNuke\end{tabular} \\ \bottomrule
		\end{tabular}
	\end{table}

	\subsection{Implementation Details}

	\textbf{1) Model Training.}  For the model architecture, the latent diffusion~\cite{Rombach_2022_CVPR} model was employed for pre-training, while the transformer-based decoder was used to aggregate features from different blocks of the UNet. The segmentation head was implemented using a traditional multilayer MLP. For collaborative learning, the feature maps of each participant were concatenated using a single-layer MLP before passing it through the segmentation head.  The model parameters for latent diffusion, ResNet34+FPN (supervised baseline), transformer-based decoder, and network in collaborative learning were 329M, 23.2M, 2M, and 361K, respectively. Latent diffusion was trained with predefined parameters from Rombach et al. ~\cite{Rombach_2022_CVPR} using a batch size of 20 and a learning rate of 2e-06. For DTSeg and collaborative learning, a batch size of 5 and the Lookahead+Radam optimizer \cite{zhang2019lookahead} with learning rates set at 1e-03 and 2e-03, respectively were used. The diffusion model adopted a similar feature extraction setup to \cite{baranchuk2021label}, utilizing 50, 150, and 200 steps for noise addition and reduction. Features were extracted from the 7th, 8th, and 9th blocks of the UNet in the latent diffusion model.  The features obtained from different steps are concatenated along the feature dimension in the same feature block. For the supervised training of both the supervised baseline and segmentation head, the Dice loss was used for training.
Additionally, conventional data augmentation strategies, such as rotation, color distortion, and scaling were used to increase the amount of data. 

    \textbf{2) Comparative Methods.} Four state-of-the-art semi-supervised semantic segmentation approaches were selected for comparison, including Adversarial Network \cite{zhang2017deep}, Cross Pseudo Supervision \cite{chen2021semi}, Uncertainty Aware Mean Teacher \cite{yu2019uncertainty}, and Deep Co-Training \cite{qiao2018deep}. It should be noted that for a fair comparison, all semi-supervised methods used the same ResNet34+FPN model as the supervised baseline.

 \subsection{Performance Metrics} 
 To evaluate the model's performance, this work employed the widely used mean intersection-over-union (mIoU) and F1 score (F1). Due to randomness and the influence of training dataset division, each experiment was repeated three times for validation, and the average and standard deviation (SD) of the results are reported. For mIoU and F1, the \textbf{highest} values are indicated in bold, while the \underline{second-highest} values are indicated with an underline. All significance tests were conducted via a t-test. In addition, results were reported for three annotation ratios: 1/20 (5\% of annotated data), 1/10 (10\% of annotated data), and 1/5 (20\% of annotated data).
	
\section{Results}	
	
	\subsection{Impact of Large-scale Diffusion Pre-training}
	Experiments were carried out on four public datasets, and the segmentation results of DTSeg are summarized in Table \ref{result_dtseg}. The dataset used for large-scale pre-training of the diffusion model included images from the PanNuke, MoNuSAC, and ConSep datasets, considering these three datasets as tests within the distribution. The CoNIC dataset was not included in the pre-training of diffusion, therefore it can be considered an OOD case for segmentation.
	
	Compared to the supervised baseline and a series of semi-supervised methods, DTSeg achieved significant improvements (p-value\textless  0.05) on both in-distribution and the OOD datasets, at different ratios of labeled datasets. Notably, the performance improvement brought by large-scale pre-training become more evident with fewer labeled images. For example, when ConSep only has 27 (1/20) labeled images, the mIoU improved from 0.413 to 0.530. Furthermore, on the MoNuSAC dataset, DTSeg outperformed the supervised baseline significantly (p-value\textless 0.05) with only 1/20 annotations. Compared to several comparative methods with 1/10 annotations, DTSeg demonstrated competitive performance, achieving a minimum improvement of 1.9\% in mIoU and 1.3\% in F1. However, when annotation levels increased to 1/5, the comparative methods showed even greater improvement than DTSeg. 
	
	
	\begin{table*}[]
		\centering
		\caption{
			Quantitative results of different methods on the PanNuke, CoNIC, MoNuSAC, and ConSep dataset. The types of methods include supervised baseline, classic semi-supervised methods, and diffusion-based segmentation methods.
		}
		\label{result_dtseg}
  \begin{threeparttable}
   \begin{adjustbox}{max width=1\textwidth,center}
		\begin{tabular}{c|cccccccccccc}
			\toprule
			\multirow{3}{*}{\textbf{PanNuke} \cite{gamper2020pannuke}}       & \multicolumn{4}{c|}{1/20 (132)}                                             & \multicolumn{4}{c|}{1/10 (265)}                                             & \multicolumn{4}{c}{1/5 (531)}                                              \\ \cline{2-13} 
			& \multicolumn{2}{c|}{\underline{mIoU}}                   & \multicolumn{2}{c|}{\underline{F1}} & \multicolumn{2}{c|}{\underline{mIoU}}                   & \multicolumn{2}{c|}{\underline{F1}} & \multicolumn{2}{c|}{\underline{mIoU}}                   & \multicolumn{2}{c}{\underline{F1}} \\
			& Mean           & \multicolumn{1}{c|}{SD}   & Mean            & \multicolumn{1}{c|}{SD}   & Mean           & \multicolumn{1}{c|}{SD}   & Mean            & \multicolumn{1}{c|}{SD}   & Mean           & \multicolumn{1}{c|}{SD}   & Mean           & SD   \\ \midrule
			Supervised Baseline \cite{lin2017feature}           & 0.420          & \multicolumn{1}{c|}{0.022} & 0.544           & \multicolumn{1}{c|}{0.028} & 0.463          & \multicolumn{1}{c|}{0.015} & 0.600           &\multicolumn{1}{c|}{0.018} & 0.492          & \multicolumn{1}{c|}{0.012} & 0.629          & 0.014 \\ \midrule
			Adversarial Network  \cite{zhang2017deep}          & 0.419          & \multicolumn{1}{c|}{0.014} & 0.539           & \multicolumn{1}{c|}{0.017} & 0.476          & \multicolumn{1}{c|}{0.010} & 0.613           & \multicolumn{1}{c|}{0.011} & \underline{0.505}          & \multicolumn{1}{c|}{0.013} & \underline{0.641}          & 0.015 \\
			Cross Pseudo Supervision \cite{chen2021semi}      & 0.420          & \multicolumn{1}{c|}{0.017} & 0.540           & \multicolumn{1}{c|}{0.021} & \underline{0.479}          & \multicolumn{1}{c|}{0.012} & \underline{0.616}           & \multicolumn{1}{c|}{0.015} & 0.503          & \multicolumn{1}{c|}{0.010} & 0.638          & 0.012 \\
			Uncertainty Aware Mean Teacher \cite{yu2019uncertainty} & 0.420          & \multicolumn{1}{c|}{0.017} & 0.539           & \multicolumn{1}{c|}{0.020} & 0.478          & \multicolumn{1}{c|}{0.011} & 0.615           & \multicolumn{1}{c|}{0.013} & 0.500          & \multicolumn{1}{c|}{0.016} & 0.636          & 0.019 \\
			Deep Co-Training   \cite{qiao2018deep}            & \underline{0.422}          & \multicolumn{1}{c|}{0.015} & \underline{0.543}           & \multicolumn{1}{c|}{0.019} & 0.476          & \multicolumn{1}{c|}{0.013} & 0.612           & \multicolumn{1}{c|}{0.017} & 0.504          & \multicolumn{1}{c|}{0.017} & 0.640          & 0.022 \\ \midrule
			DTSeg (Big)$^1$               & \textbf{0.472} & \multicolumn{1}{c|}{0.012} & \textbf{0.599}  & \multicolumn{1}{c|}{0.022} & \textbf{0.503}          & \multicolumn{1}{c|}{0.002} & \textbf{0.636}           & \multicolumn{1}{c|}{0.005} & \textbf{0.528}          & \multicolumn{1}{c|}{0.008} & \textbf{0.663}          & 0.010 \\ \midrule
			
			\midrule
			
			\multirow{3}{*}{\textbf{CoNIC} \cite{graham2021conic} (OOD case)}                                                    & \multicolumn{4}{c|}{1/20 (80)}                               & \multicolumn{4}{c|}{1/10 (160)}                              & \multicolumn{4}{c}{1/5 (320)}                               \\ \cline{2-13} 
			& \multicolumn{2}{c|}{\underline{mIoU}}          & \multicolumn{2}{c|}{\underline{F1}} & \multicolumn{2}{c|}{\underline{mIoU}}          & \multicolumn{2}{c|}{\underline{F1}} & \multicolumn{2}{c|}{\underline{mIoU}}          & \multicolumn{2}{c}{\underline{F1}} \\
			& Mean  & \multicolumn{1}{c|}{SD}   & Mean       & \multicolumn{1}{c|}{SD}        & Mean  & \multicolumn{1}{c|}{SD}   & Mean       & \multicolumn{1}{c|}{SD}        & Mean  & \multicolumn{1}{c|}{SD}   & Mean       & SD       \\ \midrule
			Supervised Baseline \cite{lin2017feature}                                                      & \underline{0.324} & \multicolumn{1}{c|}{0.008} & \underline{0.433}      & \multicolumn{1}{c|}{0.010}      & \underline{0.338} & \multicolumn{1}{c|}{0.008} & \underline{0.453}      & \multicolumn{1}{c|}{0.010}      & 0.350 & \multicolumn{1}{c|}{0.010} & \underline{0.472}      & 0.011     \\ \midrule
			Adversarial Network \cite{zhang2017deep}                                                      & 0.220 & \multicolumn{1}{c|}{0.015} & 0.308      & \multicolumn{1}{c|}{0.024}      & 0.270 & \multicolumn{1}{c|}{0.024} & 0.374      & \multicolumn{1}{c|}{0.029}      & 0.255 & \multicolumn{1}{c|}{0.007} & 0.358      & 0.016     \\
			Cross Pseudo Supervision \cite{chen2021semi}                                                  & 0.313 & \multicolumn{1}{c|}{0.014} & 0.420      & \multicolumn{1}{c|}{0.018}      & 0.322 & \multicolumn{1}{c|}{0.005} & 0.433      & \multicolumn{1}{c|}{0.006}      & 0.344 & \multicolumn{1}{c|}{0.016} & 0.462      & 0.020     \\
			Uncertainty Aware Mean Teacher \cite{yu2019uncertainty} & 0.321 & \multicolumn{1}{c|}{0.008} & 0.429      & \multicolumn{1}{c|}{0.011}      & 0.332 & \multicolumn{1}{c|}{0.012} & 0.445      & \multicolumn{1}{c|}{0.015}      & \underline{0.352} & \multicolumn{1}{c|}{0.008} & \underline{0.472}      & 0.010     \\
			Deep Co-Training \cite{qiao2018deep}                                                         & 0.315 & \multicolumn{1}{c|}{0.014} & 0.422      & \multicolumn{1}{c|}{0.019}      & 0.327 & \multicolumn{1}{c|}{0.010} & 0.438      & \multicolumn{1}{c|}{0.014}      & 0.347 & \multicolumn{1}{c|}{0.011} & 0.465      & 0.013     \\ \midrule
			DTSeg (Big)$^1$                                                               & \textbf{0.349} & \multicolumn{1}{c|}{0.010} & \textbf{0.460}      & \multicolumn{1}{c|}{0.012}      & \textbf{0.359} & \multicolumn{1}{c|}{0.009} & \textbf{0.475}      & \multicolumn{1}{c|}{0.011}      & \textbf{0.370} & \multicolumn{1}{c|}{0.007} & \textbf{0.489}      & 0.008     \\ \midrule
			
			
			\midrule
			
			\multirow{3}{*}{\textbf{MoNuSAC} \cite{verma2021monusac2020}}       & \multicolumn{4}{c|}{1/20 (70)}                                                                  & \multicolumn{4}{c|}{1/10 (140)}                                                                  & \multicolumn{4}{c}{1/5 (280)}                                              \\ \cline{2-13} 
			& \multicolumn{2}{c|}{\underline{mIoU}}                   & \multicolumn{2}{c|}{\underline{F1}}                     & \multicolumn{2}{c|}{\underline{mIoU}}                   & \multicolumn{2}{c|}{\underline{F1}}                     & \multicolumn{2}{c|}{\underline{mIoU}}                   & \multicolumn{2}{c}{\underline{F1}} \\
			& Mean           & \multicolumn{1}{c|}{SD}   & Mean           & \multicolumn{1}{c|}{SD}   & Mean           & \multicolumn{1}{c|}{SD}   & Mean           & \multicolumn{1}{c|}{SD}   & Mean           & \multicolumn{1}{c|}{SD}   & Mean           & SD   \\ \midrule
			Supervised Baseline \cite{lin2017feature}           & 0.480          & \multicolumn{1}{c|}{0.025} & 0.604          & \multicolumn{1}{c|}{0.033} & 0.517          & \multicolumn{1}{c|}{0.032} & 0.650          & \multicolumn{1}{c|}{0.031} & 0.594          & \multicolumn{1}{c|}{0.028} & \textbf{0.731}          & 0.027 \\ \midrule
			Adversarial Network \cite{zhang2017deep}           & 0.487          & \multicolumn{1}{c|}{0.008} & 0.611          & \multicolumn{1}{c|}{0.014} & \underline{0.533}          & \multicolumn{1}{c|}{0.024} & \underline{0.663}          & \multicolumn{1}{c|}{0.027} & \textbf{0.595}          & \multicolumn{1}{c|}{0.033} & 0.729          & 0.034 \\
			Cross Pseudo Supervision \cite{chen2021semi}      & \underline{0.493}          & \multicolumn{1}{c|}{0.024} & \underline{0.617}          & \multicolumn{1}{c|}{0.032} & 0.530          & \multicolumn{1}{c|}{0.031} & 0.661          & \multicolumn{1}{c|}{0.034} & \textbf{0.595}          & \multicolumn{1}{c|}{0.030} & 0.729          & 0.030 \\
			Uncertainty Aware Mean Teacher \cite{yu2019uncertainty}  & 0.487          & \multicolumn{1}{c|}{0.025} & 0.610          & \multicolumn{1}{c|}{0.033} & 0.527          & \multicolumn{1}{c|}{0.037} & 0.655          & \multicolumn{1}{c|}{0.044} & \textbf{0.595}          & \multicolumn{1}{c|}{0.033} & \underline{0.730}          & 0.034 \\
			Deep Co-Training  \cite{qiao2018deep}              & 0.485          & \multicolumn{1}{c|}{0.008} & 0.610          & \multicolumn{1}{c|}{0.016} & 0.530          & \multicolumn{1}{c|}{0.027} & 0.661          & \multicolumn{1}{c|}{0.030} & 0.594          & \multicolumn{1}{c|}{0.035} & {0.728}          & 0.037 \\ \midrule
			DTSeg (Big)$^1$                & \textbf{0.534} & \multicolumn{1}{c|}{0.019} & \textbf{0.657}          & \multicolumn{1}{c|}{0.025} & \textbf{0.552}          & \multicolumn{1}{c|}{0.027} & \textbf{0.676}          & \multicolumn{1}{c|}{0.032} & 0.583          & \multicolumn{1}{c|}{0.009} & 0.713          & 0.008 \\ 
			\midrule
			
			\midrule
			
			\multirow{3}{*}{\textbf{ConSep} \cite{graham2019hover}}  & \multicolumn{4}{c|}{1/20 (27)}                                             & \multicolumn{4}{c|}{1/10 (54)}                                             & \multicolumn{4}{c}{1/5 (108)}                                              \\ \cline{2-13} 
			& \multicolumn{2}{c|}{\underline{mIoU}}                   & \multicolumn{2}{c|}{\underline{F1}} & \multicolumn{2}{c|}{\underline{mIoU}}                   & \multicolumn{2}{c|}{\underline{F1}} & \multicolumn{2}{c|}{\underline{mIoU}}                   & \multicolumn{2}{c}{\underline{F1}} \\
			& Mean           & \multicolumn{1}{c|}{SD}   & Mean            & \multicolumn{1}{c|}{SD}   & Mean           & \multicolumn{1}{c|}{SD}   & Mean            & \multicolumn{1}{c|}{SD}   & Mean           & \multicolumn{1}{c|}{SD}   & Mean           & SD   \\ \midrule
			Supervised Baseline \cite{lin2017feature}     & 0.413          & \multicolumn{1}{c|}{0.031} & 0.530           & \multicolumn{1}{c|}{0.037} & 0.457          & \multicolumn{1}{c|}{0.006} & 0.579           & \multicolumn{1}{c|}{0.005} & 0.506          & \multicolumn{1}{c|}{0.006} & 0.634          & 0.013 \\ \midrule
			Adversarial Network \cite{zhang2017deep}   & 0.452          & \multicolumn{1}{c|}{0.018} & 0.567           & \multicolumn{1}{c|}{0.020} & \underline{0.491}          & \multicolumn{1}{c|}{0.010} & 0.609           & \multicolumn{1}{c|}{0.012} & 0.514          & \multicolumn{1}{c|}{0.005} & 0.635          & 0.006 \\
			Cross Pseudo Supervision \cite{chen2021semi} & 0.456          & \multicolumn{1}{c|}{0.016} & 0.571           & \multicolumn{1}{c|}{0.019} & 0.485          & \multicolumn{1}{c|}{0.011} & 0.603           & \multicolumn{1}{c|}{0.013} & 0.514          & \multicolumn{1}{c|}{0.002} & 0.634          & 0.004 \\
			Uncertainty Aware Mean Teacher \cite{yu2019uncertainty}             & 0.452          & \multicolumn{1}{c|}{0.010} & 0.567           & \multicolumn{1}{c|}{0.014} & 0.490          & \multicolumn{1}{c|}{0.016} & \underline{0.610}           & \multicolumn{1}{c|}{0.021} & \underline{0.515}          & \multicolumn{1}{c|}{0.002} & \underline{0.637}          & 0.004 \\
			Deep Co-Training \cite{qiao2018deep}               & \underline{0.461}          & \multicolumn{1}{c|}{0.011} & \underline{0.579}           & \multicolumn{1}{c|}{0.015} & 0.483          & \multicolumn{1}{c|}{0.006} & 0.601           & \multicolumn{1}{c|}{0.007} & 0.508          & \multicolumn{1}{c|}{0.010} & 0.628          & 0.012 \\ \midrule
			DTSeg (Big)$^1$          & \textbf{0.530} & \multicolumn{1}{c|}{0.014} & \textbf{0.657}  & \multicolumn{1}{c|}{0.024} & \textbf{0.551} & \multicolumn{1}{c|}{0.014} & \textbf{0.676}  & \multicolumn{1}{c|}{0.021} & \textbf{0.568} & \multicolumn{1}{c|}{0.012} & \textbf{0.691} & 0.016 \\ \bottomrule
			
		\end{tabular}
  \end{adjustbox}
  \begin{tablenotes}
      \item[1] As shown in Table \ref{tab:pretrained}, the explanation for the pre-training dataset indicates that DTSeg (Big) represents the utilization  \\ of pre-trained diffusion (Big).
    \end{tablenotes}
  \end{threeparttable}
	\end{table*}

	\begin{table*}[]
		\centering
		\caption{
			Quantitative results of the collaborative learning on the PanNuke, CoNIC, and MoNuSAC dataset. The types of methods include supervised baseline, diffusion-based segmentation methods, and collaborative learning. 
		}
		\label{tab:collaborative learning}
  \begin{threeparttable}
  \begin{adjustbox}{max width=1\textwidth,center}
		\begin{tabular}{c|cccccccccccc}
			\toprule
			
			\multirow{3}{*}{\textbf{PanNuke} \cite{gamper2020pannuke}}       & \multicolumn{4}{c|}{1/20 (132)}                                             & \multicolumn{4}{c|}{1/10 (265)}                                             & \multicolumn{4}{c}{1/5 (531)}                                              \\ \cline{2-13} 
			& \multicolumn{2}{c|}{\underline{mIoU}}                   & \multicolumn{2}{c|}{\underline{F1}} & \multicolumn{2}{c|}{\underline{mIoU}}                   & \multicolumn{2}{c|}{\underline{F1}} & \multicolumn{2}{c|}{\underline{mIoU}}                   & \multicolumn{2}{c}{\underline{F1}} \\
			& Mean           & \multicolumn{1}{c|}{SD}   & Mean            & \multicolumn{1}{c|}{SD}    & Mean           & \multicolumn{1}{c|}{SD}   & Mean            & \multicolumn{1}{c|}{SD}    & Mean           & \multicolumn{1}{c|}{SD}   & Mean           & SD   \\ \midrule
			Supervised Baseline \cite{lin2017feature}           & 0.420          & \multicolumn{1}{c|}{0.022} & 0.544           & \multicolumn{1}{c|}{0.028} & 0.463          & \multicolumn{1}{c|}{0.015} & 0.600           & \multicolumn{1}{c|}{0.018} & 0.492          & \multicolumn{1}{c|}{0.012} & 0.629          & 0.014 \\ \midrule
			DTSeg (MoNuSAC)            &0.422            & \multicolumn{1}{c|}{0.020}   & 0.544             & \multicolumn{1}{c|}{0.028}   
			& 0.448            & \multicolumn{1}{c|}{0.003}   &
			0.576            & \multicolumn{1}{c|}{0.005}    & 0.474            & \multicolumn{1}{c|}{0.012}   & 0.607           & 0.014   \\
			DTSeg (PanNuke)            & {0.447}          & \multicolumn{1}{c|}{0.019} & {0.571}           & \multicolumn{1}{c|}{0.026} & 0.469          & \multicolumn{1}{c|}{0.029} & 0.598           & \multicolumn{1}{c|}{0.040} & 0.504          & \multicolumn{1}{c|}{0.018} & 0.638          & 0.021 \\ 
			DTSeg (Big)                & \textbf{0.472} & \multicolumn{1}{c|}{0.012} & \textbf{0.599}  & \multicolumn{1}{c|}{0.022} & \underline{0.503}          & \multicolumn{1}{c|}{0.002} & \underline{0.636}           & \multicolumn{1}{c|}{0.005} & \underline{0.528}          & \multicolumn{1}{c|}{0.008} & \underline{0.663}          & 0.010 \\ \midrule
			Collaboration (MoNuSAC)        & 0.432            & \multicolumn{1}{c|}{0.024}   & 0.554             &\multicolumn{1}{c|}{0.029}   
			& {0.489}   & \multicolumn{1}{c|}{0.009}   & 	 
			{0.626}    & \multicolumn{1}{c|}{0.011}    & {0.513}            & \multicolumn{1}{c|}{0.010 }   & 0.649            & 0.011   \\
			Collaboration (PanNuke)        & {0.453}          & \multicolumn{1}{c|}{0.008} & {0.576}           & \multicolumn{1}{c|}{0.015} & {0.494}          & \multicolumn{1}{c|}{0.022} & {0.632}           & \multicolumn{1}{c|}{0.023} & {0.526}          & \multicolumn{1}{c|}{0.013} & 0.662          & 0.015 \\
			Collaboration (Big)            & \underline{0.468}          & \multicolumn{1}{c|}{0.019} & \underline{0.594}           & \multicolumn{1}{c|}{0.026} & \textbf{0.505} & \multicolumn{1}{c|}{0.006} & \textbf{0.641}  & \multicolumn{1}{c|}{0.008} & \textbf{0.530} & \multicolumn{1}{c|}{0.013} & \textbf{0.665} & 0.013 \\ 
			\midrule \midrule
			
			\multirow{3}{*}{\textbf{CoNIC} \cite{graham2021conic} (OOD case)}                                                    & \multicolumn{4}{c|}{1/20 (80)}                               & \multicolumn{4}{c|}{1/10 (160)}                              & \multicolumn{4}{c}{1/5 (320)}                               \\ \cline{2-13} 
			& \multicolumn{2}{c|}{\underline{mIoU}}          & \multicolumn{2}{c|}{\underline{F1}} & \multicolumn{2}{c|}{\underline{mIoU}}          & \multicolumn{2}{c|}{\underline{F1}} & \multicolumn{2}{c|}{\underline{mIoU}}          & \multicolumn{2}{c}{\underline{F1}} \\
			& Mean  & \multicolumn{1}{c|}{SD}   & Mean       & \multicolumn{1}{c|}{SD}        & Mean  & \multicolumn{1}{c|}{SD}   & Mean       & \multicolumn{1}{c|}{SD}        & Mean  & \multicolumn{1}{c|}{SD}   & Mean       & SD       \\ \midrule
			Supervised Baseline \cite{lin2017feature}                                                      & 0.324 & \multicolumn{1}{c|}{0.008} & 0.433      & \multicolumn{1}{c|}{0.010}      & 0.338 & \multicolumn{1}{c|}{0.008} & 0.453      & \multicolumn{1}{c|}{0.010}      & 0.350 & \multicolumn{1}{c|}{0.010} & 0.472      & 0.011     \\ \midrule
			DTSeg (MoNuSAC)                                                           & 0.309 & \multicolumn{1}{c|}{0.008} & 0.414      & \multicolumn{1}{c|}{0.009}      & 0.341 & \multicolumn{1}{c|}{0.005} & 0.456      & \multicolumn{1}{c|}{0.004}      & 0.361 & \multicolumn{1}{c|}{0.002} & 0.481      & 0.004     \\
			DTSeg (PanNuke)                                                           & 0.337 & \multicolumn{1}{c|}{0.015} & 0.447      & \multicolumn{1}{c|}{0.020}      & 0.345 & \multicolumn{1}{c|}{0.002} & 0.461      & \multicolumn{1}{c|}{0.004}      & 0.365 & \multicolumn{1}{c|}{0.007} & 0.485      & 0.010     \\
			DTSeg (Big)                                                               & \textbf{0.349} & \multicolumn{1}{c|}{0.010} & \textbf{0.460}      & \multicolumn{1}{c|}{0.012}      & 0.359 & \multicolumn{1}{c|}{0.009} & 0.475      & \multicolumn{1}{c|}{0.011}      & 0.370 & \multicolumn{1}{c|}{0.007} & 0.489      & 0.008     \\ \midrule
			Collaboration (MoNuSAC)                                                   & 0.345 & \multicolumn{1}{c|}{0.003} & 0.456      & \multicolumn{1}{c|}{0.005}      & 0.362 & \multicolumn{1}{c|}{0.004} & 0.481      & \multicolumn{1}{c|}{0.006}      & 0.374 & \multicolumn{1}{c|}{0.008} & 0.500      & 0.009     \\
			Collaboration (PanNuke)                                                   & 0.345 & \multicolumn{1}{c|}{0.009} & 0.458      & \multicolumn{1}{c|}{0.011}      & \underline{0.364} & \multicolumn{1}{c|}{0.013} & \underline{0.482}      & \multicolumn{1}{c|}{0.017}      & \underline{0.383} & \multicolumn{1}{c|}{0.012} & \underline{0.508}      & 0.013     \\
			Collaboration (Big)                                                       & \underline{0.348} & \multicolumn{1}{c|}{0.025} & \textbf{0.460 }     & \multicolumn{1}{c|}{0.031}      & \textbf{0.370} & \multicolumn{1}{c|}{0.007} & \textbf{0.487}      & \multicolumn{1}{c|}{0.010}      & \textbf{0.393} & \multicolumn{1}{c|}{0.005} & \textbf{0.520}      & 0.004     \\  \midrule \midrule
			
			\multirow{3}{*}{\textbf{MoNuSAC} \cite{verma2021monusac2020}}       & \multicolumn{4}{c|}{1/20 (70)}                                                                  & \multicolumn{4}{c|}{1/10 (140)}                                                                  & \multicolumn{4}{c}{1/5 (280)}                                              \\ \cline{2-13} 
			& \multicolumn{2}{c|}{\underline{mIoU}}                   & \multicolumn{2}{c|}{\underline{F1}}                     & \multicolumn{2}{c|}{\underline{mIoU}}                   & \multicolumn{2}{c|}{\underline{F1}}                     & \multicolumn{2}{c|}{\underline{mIoU}}                   & \multicolumn{2}{c}{\underline{F1}} \\
			& Mean           & \multicolumn{1}{c|}{SD}   & Mean           & \multicolumn{1}{c|}{SD}   & Mean           & \multicolumn{1}{c|}{SD}   & Mean           & \multicolumn{1}{c|}{SD}   & Mean           & \multicolumn{1}{c|}{SD}   & Mean           & SD   \\ \midrule
			Supervised Baseline \cite{lin2017feature}           & 0.480          & \multicolumn{1}{c|}{0.025} & 0.604          & \multicolumn{1}{c|}{0.033} & 0.517          & \multicolumn{1}{c|}{0.032} & 0.650          & \multicolumn{1}{c|}{0.031} & 0.594          & \multicolumn{1}{c|}{0.028} & 0.731          & 0.027 \\ \midrule
			DTSeg (MoNuSAC)            & {0.521}          & \multicolumn{1}{c|}{0.029} & {0.641}          & \multicolumn{1}{c|}{0.036} & 0.539          & \multicolumn{1}{c|}{0.021} & 0.661          & \multicolumn{1}{c|}{0.027} & 0.562          & \multicolumn{1}{c|}{0.009} & 0.690          & 0.013 \\
			DTSeg (PanNuke)            & 0.500          & \multicolumn{1}{c|}{0.008} & 0.618          & \multicolumn{1}{c|}{0.014} & 0.538          & \multicolumn{1}{c|}{0.020} & 0.661          & \multicolumn{1}{c|}{0.022} & 0.566          & \multicolumn{1}{c|}{0.016} & 0.695          & 0.016 \\
			DTSeg (Big)                & \textbf{0.534} & \multicolumn{1}{c|}{0.019} & \underline{0.657}          & \multicolumn{1}{c|}{0.025} & {0.552}          & \multicolumn{1}{c|}{0.027} & {0.676}          & \multicolumn{1}{c|}{0.032} & 0.583          & \multicolumn{1}{c|}{0.009} & 0.713          & 0.008 \\ 
			\midrule
			Collaboration (MoNuSAC)        & {0.523}          & \multicolumn{1}{c|}{0.027} & {0.645}          & \multicolumn{1}{c|}{0.030} & \textbf{0.562} & \multicolumn{1}{c|}{0.016} & \textbf{0.688} & \multicolumn{1}{c|}{0.019} & \underline{0.612}          & \multicolumn{1}{c|}{0.027} & \underline{0.746}          & 0.024 \\
			Collaboration (PanNuke)        & 0.496          & \multicolumn{1}{c|}{0.027} & 0.624          & \multicolumn{1}{c|}{0.029} & {0.543}          & \multicolumn{1}{c|}{0.018} & {0.674}          & \multicolumn{1}{c|}{0.027} & {0.596}          & \multicolumn{1}{c|}{0.020} & 0.733          & 0.020 \\
			Collaboration (Big)            & \underline{0.533}          & \multicolumn{1}{c|}{0.016} & \textbf{0.660} & \multicolumn{1}{c|}{0.023} & \underline{0.560}         & \multicolumn{1}{c|}{0.014} & \underline{0.687}          & \multicolumn{1}{c|}{0.018} & \textbf{0.622} & \multicolumn{1}{c|}{0.016} & \textbf{0.754} & 0.014 \\ 
			\bottomrule
		\end{tabular}
    \end{adjustbox}
    \begin{tablenotes}
      \item[1] As shown in Table \ref{tab:pretrained}, we reported results for three different pre-training weights: MoNuSAC, PanNuke, and Big.
    \end{tablenotes}
  \end{threeparttable}
	\end{table*}

To further investigate the superiority of pre-trained diffusion, we have a more in-depth discussion (Fig. \ref{supervised_100}, Table \ref{simsiam} and Fig. \ref{figure_umap}). Our findings are summarized below.
\subsubsection{Comparison with Supervised Baseline}
The DTSeg model surpassed the performance of fully labeled supervised baselines using only a small fraction of the available labels. In the case of the ConSep dataset (Fig. \ref{supervised_100}a), utilizing only 10\% of the available labels DTSeg achieved comparable performance to the supervised method trained with 100\% labeled data. 

\subsubsection{Comparison with SimSiam} As shown in Table \ref{simsiam}, it was found that SimSiam, typically utilized for pre-training with unlabeled data in classification tasks, did not achieve comparable performance with DTSeg.  This suggests that conventional pre-training-based methods that are applied for classification tasks, may not learn semantic information during pre-training.  For MoNuSAC dataset, the DTSeg model outperformed the results achieved by SimSiam pre-training.

\subsubsection{Feature Clustering using UMAP}
\begin{figure}[h]
		\centerline{\includegraphics[width=\linewidth]{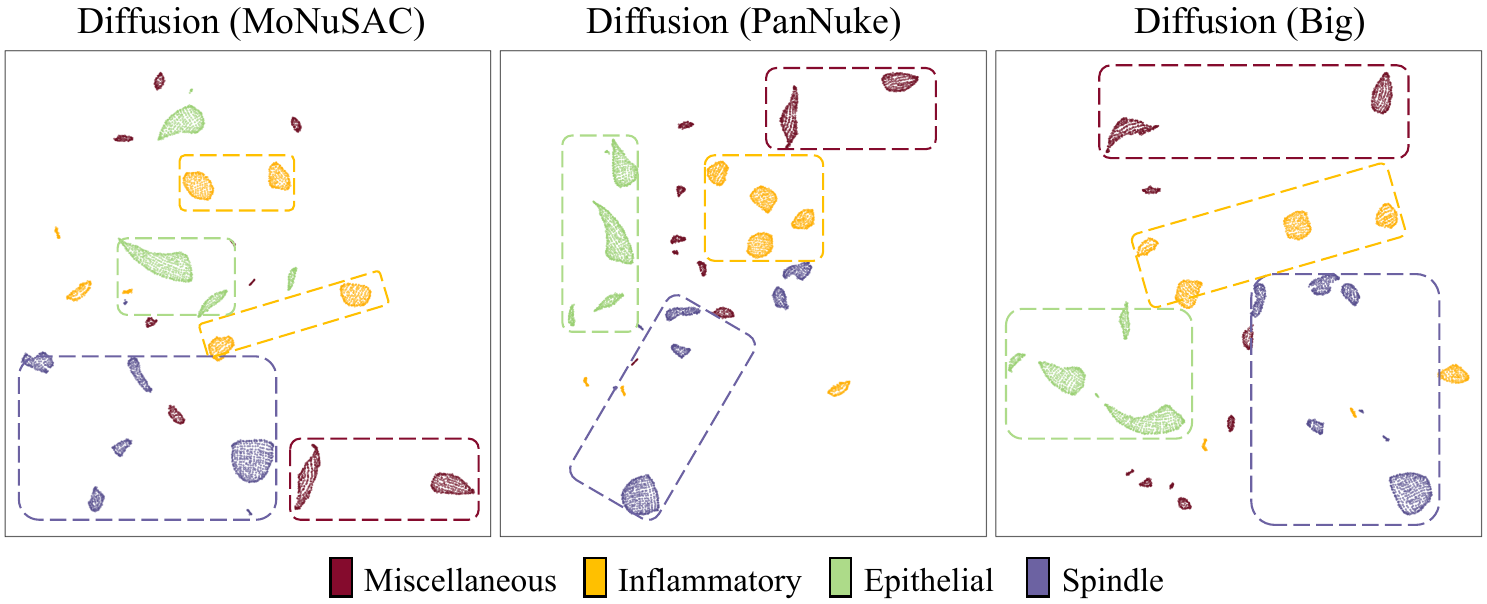}}
		\caption{\textbf{Visualization of features extracted by pre-trained diffusion models.} We used the ConSep dataset and UMAP~\cite{mcinnes2018umap-software} to visualize the features extracted by pre-trained diffusion models. Specifically, we selected 1000 pixels for each category.
		}
		\label{figure_umap}
	\end{figure}

The effective clustering of three pre-trained weights using Uniform Manifold Approximation and Projection (UMAP) \cite{mcinnes2018umap} is demonstrated in Fig.~\ref{figure_umap}. Different colors represent various types of cell nuclei, with the dashed box used to emphasize the clustering in the feature space. In summary, our findings indicate that the categories exhibit a close distribution in the high-dimensional feature space, suggesting a distinct feature mapping for different cell nucleus categories.

	\begin{figure}[htbp]
		\centering
		
		\subfigure[ConSep.]{
			\includegraphics[width=0.46\linewidth]{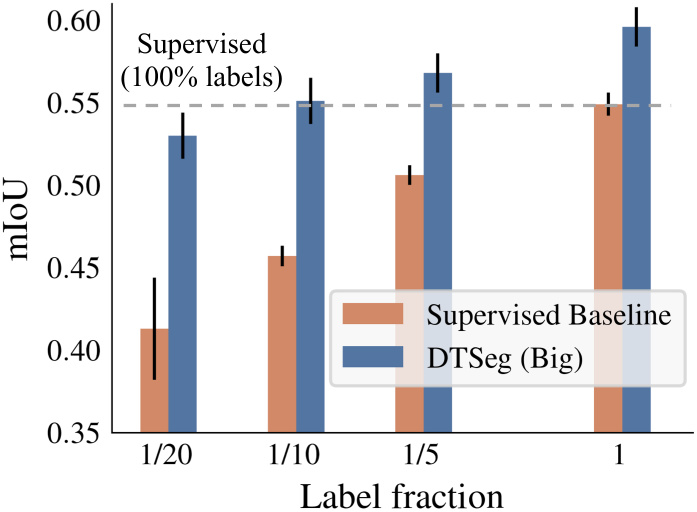}
		}
		\subfigure[MoNuSAC. ]{
			\includegraphics[width=0.46\linewidth]{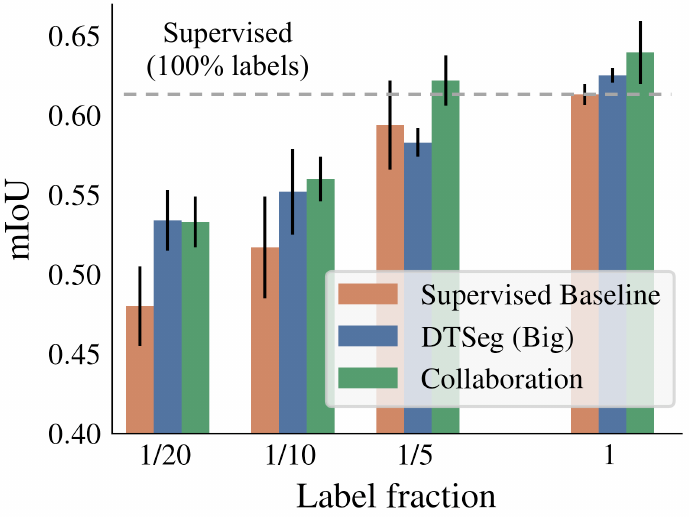}
		}
		\caption{ \textbf{Comparison with Supervised Baseline using 100\% Labels.} Diffusion-based large-scale pre-training can achieve comparable performance with supervised baselines. Collaborative learning can help further improve performance. }
    \label{supervised_100}
	\end{figure}

	\begin{table}[]
		\centering
		\caption{
			Comparison of Different Pre-Training Approaches.
		}
		\label{simsiam}
            \setlength{\tabcolsep}{1.8mm}
            \begin{threeparttable}
            \scalebox{1.0}{
		\begin{tabular}{c|cc|cc|cc}
			\toprule
			\multirow{2}{*}{\textbf{MoNuSAC}\cite{verma2021monusac2020}} & \multicolumn{2}{c|}{\underline{1/20 (70)}}                               & \multicolumn{2}{c|}{\underline{1/10 (140)}}                              & \multicolumn{2}{c}{\underline{1/5 (280)}}                               \\ 
			& Mean  & \multicolumn{1}{c|}{SD}     & Mean  & \multicolumn{1}{c|}{SD}          & Mean  & \multicolumn{1}{c}{SD}       \\ \midrule
			Supervised \cite{lin2017feature}      & \underline{0.480} & \multicolumn{1}{c|}{0.025}   & \underline{0.517} & \multicolumn{1}{c|}{0.032}      & \textbf{0.594} & \multicolumn{1}{c}{0.028}     \\
			SimSiam (Big)  \cite{chen2021exploring}               & 0.400 & \multicolumn{1}{c|}{0.016}   & 0.422 & \multicolumn{1}{c|}{0.022}       & 0.471 & \multicolumn{1}{c}{0.009}     \\
			DTSeg (Big)  \cite{Rombach_2022_CVPR}             & \textbf{0.534} & \multicolumn{1}{c|}{0.019}  & \textbf{0.552} & \multicolumn{1}{c|}{0.027}    & \underline{0.583} & \multicolumn{1}{c}{0.009}  \\ \bottomrule
		\end{tabular}
            }
            \end{threeparttable}
	\end{table}

\subsection{Impact of Collaborative Learning}
The effect of collaborative learning was explored with multiple datasets.  Table \ref{tab:collaborative learning} shows the results of this investigation on two aspects: \textbf{1)} the effectiveness of collaborative learning with limited pre-training data, and \textbf{2)} the feasibility of applying collaborative learning to both in-distribution and OOD segmentation datasets. 
	
	When DTSeg was pre-trained using small datasets like MoNuSAC and PanNuke, collaborative learning was found to significantly enhance its performance. As shown in Table \ref{tab:collaborative learning}, both Collaboration (MoNuSAC) and Collaboration (PanNuke) consistently outperformed DTSeg (MoNuSAC) and DTSeg (PanNuke) respectively as the number of labeled images increased (p-value\textless 0.05). Additionally, when the models were tested on a dataset that was not included in the pre-training, such as CoNIC, collaborative learning also demonstrated the potential to improve the performance of DTSeg with an increasing number of labeled images.
Furthermore, compared with fully labeled supervised baselines, collaborative learning showed superior performance on the MoNuSAC dataset (Fig. \ref{supervised_100}b), even when utilizing only 20\% of the available labels.

	\subsection{Visualization of Segmentation Results}

	\begin{figure*}[!t]
		\centerline{\includegraphics[width=\linewidth]{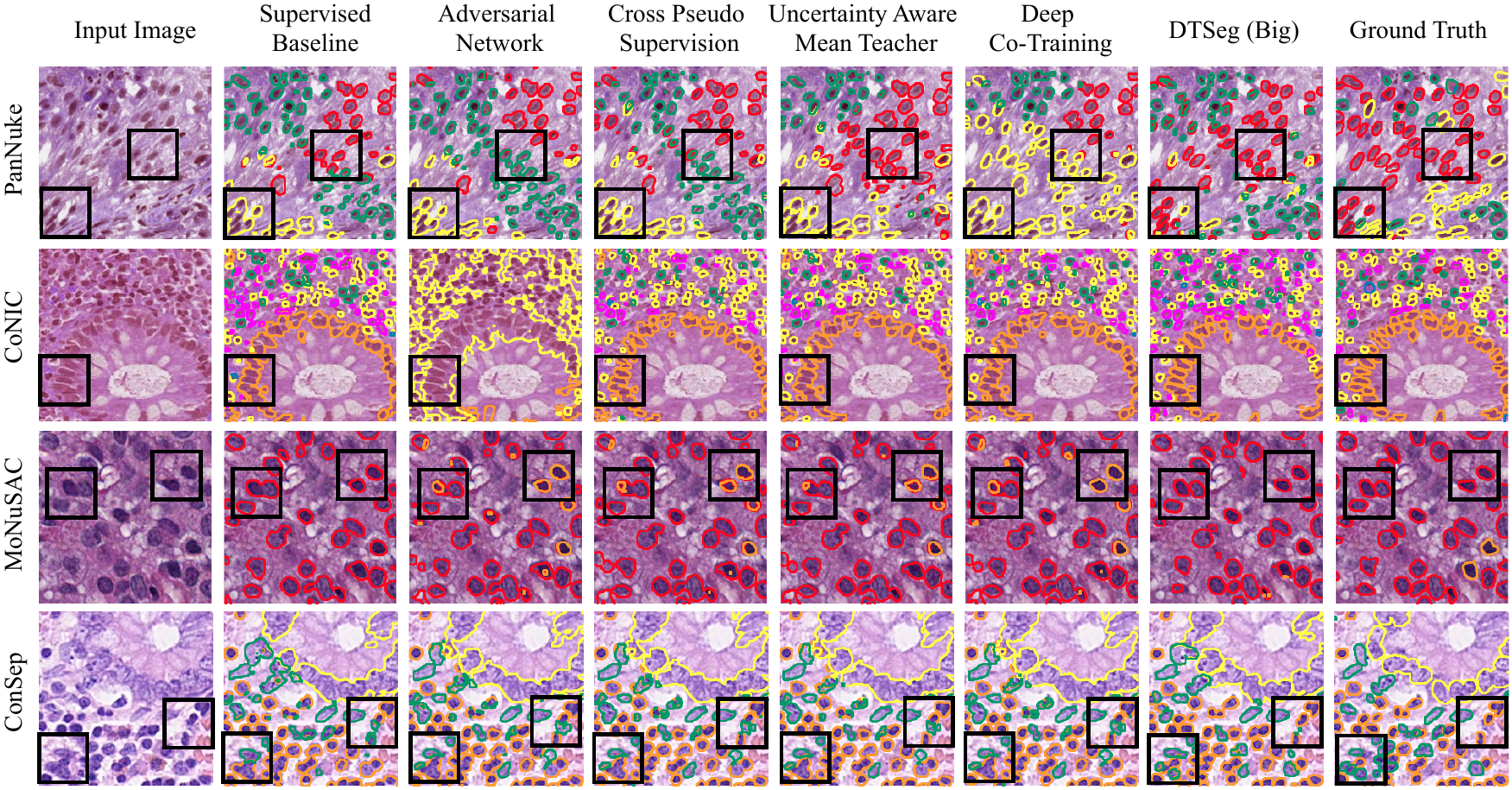}}
		\caption{\textbf{Visualization of semantic segmentation results for cell nuclei.} We visualized the segmentation results of different methods on different datasets. The reference table for the different colors corresponding to cell nuclear categories is provided in Fig. \ref{figure_visualization_dataset}.
		}
		\label{figure_visualization_predict}
	\end{figure*}

	\begin{figure}[!t]
		\centerline{\includegraphics[width=\linewidth]{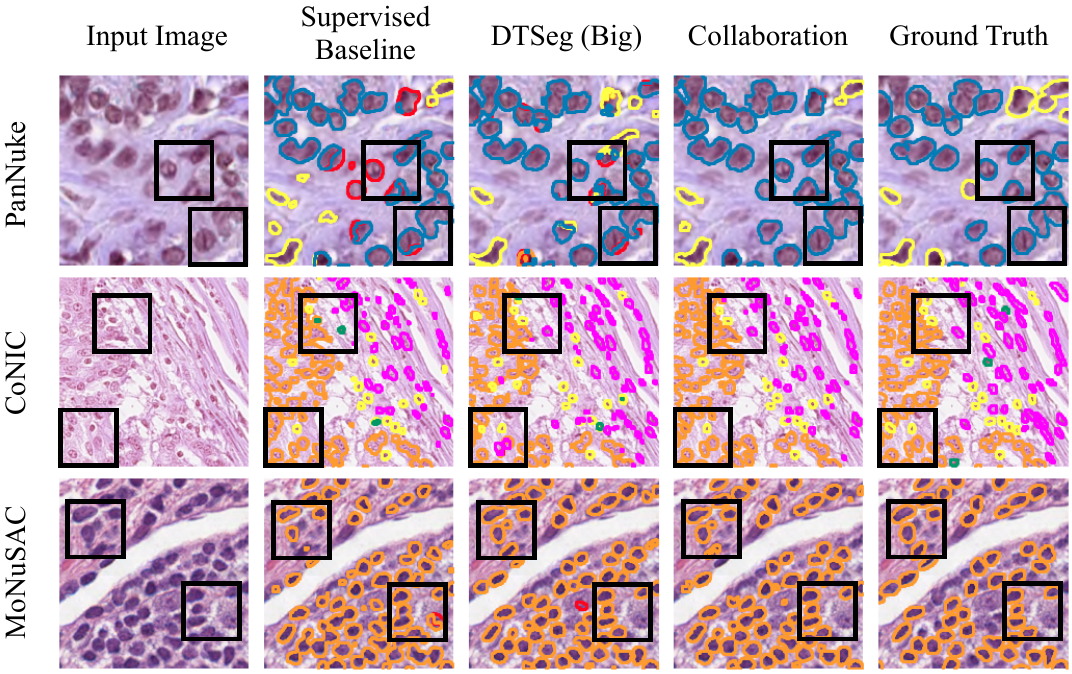}}
		\caption{\textbf{Visualization of collaborative learning results for OOD dataset.} Specifically, we visualized the predictions of Collaboration (MoNuSAC) on the PanNuke dataset, the predictions of Collaboration (Big) on the CoNIC dataset, and the predictions of Collaboration (PanNuke) on the MoNuSAC dataset.
		}
		\label{figure_visualization_predict_ood}
	\end{figure}

 Figure \ref{figure_visualization_predict} shows semantic results of DTSeg (Big) with other competing semi-supervised methods from each dataset. The black bounding boxes highlight the regions where DTSeg (Big) outperformed other methods in cell nuclei segmentation across diverse datasets. Results of collaborative learning are shown in Fig. \ref{figure_visualization_predict_ood} and the performance improvement over DTSeg (Big) can be observed within the black bounding boxes.  
	
     
	\begin{table}[]
		\centering
		\caption{
			Impact of Self-Attention (SA) in Transformer Decoder.
		}
		\label{ablation_sa}
            \setlength{\tabcolsep}{1.8mm}
        \begin{threeparttable}
		\begin{tabular}{c|cc|cc|cc}
			\toprule
			\multirow{2}{*}{\textbf{MoNuSAC}\cite{verma2021monusac2020}} & \multicolumn{2}{c|}{\underline{1/20 (70)}}                                        & \multicolumn{2}{c|}{\underline{1/10 (140)}}                                       & \multicolumn{2}{c}{\underline{1/5 (280)}}                                        \\ 
			& Mean           & \multicolumn{1}{c|}{SD}    & Mean           & \multicolumn{1}{c|}{SD}    & Mean           & \multicolumn{1}{c}{SD}     \\ \midrule
			w/o SA (MoNuSAC)      & 0.510          & \multicolumn{1}{c|}{0.016}  & 0.531          & \multicolumn{1}{c|}{0.009} & 0.555          & \multicolumn{1}{c}{0.009}  \\
			w/ SA (MoNuSAC)       & \textbf{0.521} & \multicolumn{1}{c|}{0.029}  & \textbf{0.539} & \multicolumn{1}{c|}{0.021}  & \textbf{0.562} & \multicolumn{1}{c}{0.009}  \\ \midrule
			w/o SA (PanNuke)      & 0.489          & \multicolumn{1}{c|}{0.013}  & 0.508          & \multicolumn{1}{c|}{0.009} & 0.532          & \multicolumn{1}{c}{0.009} \\
			w/ SA (PanNuke)       & \textbf{0.500} & \multicolumn{1}{c|}{0.008} & \textbf{0.538} & \multicolumn{1}{c|}{0.020} & \textbf{0.566} & \multicolumn{1}{c}{0.016} \\ \midrule
			w/o SA (Big)          & \textbf{0.538}          & \multicolumn{1}{c|}{0.010} & {0.542} & \multicolumn{1}{c|}{0.011} & {0.580} & \multicolumn{1}{c}{0.011}  \\
			w/ SA (Big)           & {0.534} & \multicolumn{1}{c|}{0.019} & \textbf{0.552}          & \multicolumn{1}{c|}{0.027}  & \textbf{0.583}          & \multicolumn{1}{c}{0.009}  \\ \bottomrule
		\end{tabular}
      \begin{tablenotes}
      \item[1] Information on three different pre-training diffusion models (MoNuSAC, PanNuke, and Big) is presented in Table \ref{tab:pretrained}.
    \end{tablenotes}
  \end{threeparttable}
	\end{table}

	\begin{table}[h]
		\centering
		\caption{
			Serial vs. Parallel Processing of Feature Blocks on diffusion-based Semantic Segmentation Framework.
		}
		\label{ablation_feature_blocks}
  \begin{threeparttable}
            \setlength{\tabcolsep}{1.8mm}
		\begin{tabular}{c|cc|cc|cc}
			\toprule
			\multirow{2}{*}{\textbf{MoNuSAC}\cite{verma2021monusac2020}} & \multicolumn{2}{c|}{\underline{1/20 (70)}}                                        & \multicolumn{2}{c|}{\underline{1/10 (140)}}                                       & \multicolumn{2}{c}{\underline{1/5 (280)}}                                    \\
			& Mean           & \multicolumn{1}{c|}{SD}      & Mean           & \multicolumn{1}{c|}{SD}     & Mean           & \multicolumn{1}{c}{SD}    \\ \midrule
			Serial (MoNuSAC)         & 0.503          & \multicolumn{1}{c|}{0.019} & 0.532          & \multicolumn{1}{c|}{0.007}  & 0.559          & \multicolumn{1}{c}{0.020}  \\
			Parallel (MoNuSAC)       & \textbf{0.521} & \multicolumn{1}{c|}{0.029}  & \textbf{0.539} & \multicolumn{1}{c|}{0.021}  & \textbf{0.562} & \multicolumn{1}{c}{0.009} \\ \midrule
			Serial (PanNuke)         & 0.483          & \multicolumn{1}{c|}{0.016}  & 0.532          & \multicolumn{1}{c|}{0.006}  & 0.557          & \multicolumn{1}{c}{0.009}  \\
			Parallel (PanNuke)       & \textbf{0.500} & \multicolumn{1}{c|}{0.008}  & \textbf{0.538} & \multicolumn{1}{c|}{0.020}  & \textbf{0.566} & \multicolumn{1}{c}{0.016} \\ \midrule
			Serial (Big)             & 0.532          & \multicolumn{1}{c|}{0.017}  & \textbf{0.554} & \multicolumn{1}{c|}{0.010}  & \textbf{0.601} & \multicolumn{1}{c}{0.018}  \\
			Parallel (Big)           & \textbf{0.534} & \multicolumn{1}{c|}{0.019}  & 0.552          & \multicolumn{1}{c|}{0.027}  & 0.583          & \multicolumn{1}{c}{0.009} \\ \bottomrule
		\end{tabular}
        \begin{tablenotes}
      \item[1] Information on three different pre-training diffusion models (MoNuSAC, PanNuke, and Big) is presented in Table \ref{tab:pretrained}.
    \end{tablenotes}
  \end{threeparttable}
	\end{table}

	
	\begin{table}[]
		\centering
		\caption{
			Collaborative Learning Comparison between DTSeg and Traditional Semantic Segmentation.
		}
		\label{ablation_collaboration}
  \begin{threeparttable}
            \setlength{\tabcolsep}{1.8mm}
		\begin{tabular}{c|cc|cc|cc}
			\toprule
			\multirow{2}{*}{\textbf{MoNuSAC} \cite{verma2021monusac2020}}            & \multicolumn{2}{c|}{\underline{1/20 (70)}}                                        & \multicolumn{2}{c|}{\underline{1/10 (140)}}                                       & \multicolumn{2}{c}{\underline{1/5 (280)}}                                         \\
			& Mean           & \multicolumn{1}{c|}{SD}   & Mean                     & \multicolumn{1}{c|}{SD}   & Mean                & \multicolumn{1}{c}{SD}   \\ \midrule
			\textbf{ResNet34+FPN} $^1$                        & 0.480          & \multicolumn{1}{c|}{0.025} & 0.517          & \multicolumn{1}{c|}{0.032}  & \underline{0.594}          & \multicolumn{1}{c}{0.028} \\
			DTSeg (Big)                     & \textbf{0.534} & \multicolumn{1}{c|}{0.019} & \underline{0.552}          & \multicolumn{1}{c|}{0.027}  & 0.583          & \multicolumn{1}{c}{0.009}  \\ \midrule
			FPN+FPN             & 0.479          & \multicolumn{1}{c|}{0.030}  & 0.480          & \multicolumn{1}{c|}{0.027}  & 0.589          & \multicolumn{1}{c}{0.034} \\
			FPN+UNet $^2$           & 0.383          & \multicolumn{1}{c|}{0.023}  & 0.518          & \multicolumn{1}{c|}{0.019}  & 0.581          & \multicolumn{1}{c}{0.024}  \\
			FPN+PSPNet          & 0.484          & \multicolumn{1}{c|}{0.015}  & 0.511          & \multicolumn{1}{c|}{0.025}  & 0.570          & \multicolumn{1}{c}{0.055}  \\
			Diffusion+Diffusion & 0.482          & \multicolumn{1}{c|}{0.085}  & 0.544          & \multicolumn{1}{c|}{0.018}  & 0.573          & \multicolumn{1}{c}{0.010}  \\
			FPN+Diffusion       & \underline{0.533}          & \multicolumn{1}{c}{0.016}  & \textbf{0.560} & \multicolumn{1}{c|}{0.014}  & \textbf{0.622} & \multicolumn{1}{c}{0.016}  \\ \midrule \midrule
			\textbf{EfficientNet+UNet}                   & 0.439          & \multicolumn{1}{c|}{0.012}  & 0.509          & \multicolumn{1}{c|}{0.040}  & 0.601          & \multicolumn{1}{c}{0.035}  \\
			DTSeg (Big)                     & \textbf{0.534} & \multicolumn{1}{c|}{0.019}  & \underline{0.552}          & \multicolumn{1}{c|}{0.027}  & 0.583          & \multicolumn{1}{c}{0.009} \\ \midrule
			UNet+UNet        & 0.443          & \multicolumn{1}{c|}{0.017}  & 0.523          & \multicolumn{1}{c|}{0.023}  & \underline{0.604}          & \multicolumn{1}{c}{0.026}  \\
			UNet+FPN            & 0.461          & \multicolumn{1}{c|}{0.026}  & 0.508          & \multicolumn{1}{c|}{0.023}  & 0.566          & \multicolumn{1}{c}{0.022}  \\
			UNet+PSPNet         & 0.457          & \multicolumn{1}{c|}{0.004}  & 0.518          & \multicolumn{1}{c|}{0.023}  & 0.603          & \multicolumn{1}{c}{0.029} \\
			Diffusion+Diffusion & \underline{0.482}          & \multicolumn{1}{c|}{0.085}  & 0.544          & \multicolumn{1}{c|}{0.018}  & 0.573          & \multicolumn{1}{c}{0.010}  \\
			UNet+Diffusion      & 0.445          & \multicolumn{1}{c|}{0.043}  & \textbf{0.556} & \multicolumn{1}{c|}{0.017}  & \textbf{0.611} & \multicolumn{1}{c}{0.011} \\ \bottomrule
		\end{tabular}
          \begin{tablenotes}
      \item[1] ResNet34 serves as the encoder, while FPN functions as the decoder.
      \item[2] ResNet34 serves as the encoder in both cases, with FPN and UNet acting as two decoders for participants in collaborative learning.
    \end{tablenotes}
  \end{threeparttable}
	\end{table}

	\begin{table}[]
		\centering
		\caption{
			Exploring Collaborative Learning Effects on Semi-supervised Methods.
		}
		\label{collaboration_semi}
  \begin{threeparttable}
		\begin{tabular}{c|cc|cc|cc}
			\toprule
			\multirow{2}{*}{\textbf{MoNuSAC}\cite{verma2021monusac2020}}                                                                 & \multicolumn{2}{c|}{\underline{1/20 (70)}}                               & \multicolumn{2}{c|}{\underline{1/10 (140)}}                              & \multicolumn{2}{c}{\underline{1/5 (280)}}                              \\
			& Mean  & \multicolumn{1}{c|}{SD}       & Mean  & \multicolumn{1}{c|}{SD}          & Mean  & \multicolumn{1}{c}{SD}         \\ \midrule
			{AN} \cite{zhang2017deep}            & 0.465 & \multicolumn{1}{c|}{0.091}   & 0.551 & \multicolumn{1}{c|}{0.008}      & 0.606 & \multicolumn{1}{c}{0.031}    \\
			 CPS \cite{chen2021semi}        & \underline{0.518} & \multicolumn{1}{c|}{0.024}   & \underline{0.558} & \multicolumn{1}{c|}{0.012}       & 0.607 & \multicolumn{1}{c}{0.015}      \\
    UAMT \cite{yu2019uncertainty}  & 0.485 & \multicolumn{1}{c|}{0.057}   & 0.537 & \multicolumn{1}{c|}{0.014}       & \underline{0.614} & \multicolumn{1}{c}{0.011}      \\
		DCT \cite{qiao2018deep}                & 0.390 & \multicolumn{1}{c|}{0.087} & 0.526 & \multicolumn{1}{c|}{0.032}   & 0.606 & \multicolumn{1}{c}{0.030}   \\ \midrule
	{Supervised} \cite{lin2017feature}           & \textbf{0.539} & \multicolumn{1}{c|}{0.023}  & \textbf{0.560} & \multicolumn{1}{c|}{0.014}    & \textbf{0.622} & \multicolumn{1}{c}{0.016}     \\ \bottomrule
		\end{tabular}
            \begin{tablenotes}
      \item[1] AN, CPS, UAMT and DCT are four different semi-supervised methods.
    \end{tablenotes}
  \end{threeparttable}
	\end{table}

	\subsection{Ablation Study}
	Several ablation studies were performed to analyse the impact of different components of the proposed framework. The MoNuSAC dataset served as the basis for all ablation experiments. Mean and standard deviation OF mIOU scores were computed as performance measures for all the ablation studies over three distinct data partitions. For Table~\ref{ablation_sa} and \ref{ablation_feature_blocks}, we reported results for three different pre-training weights (Details in Table \ref{tab:pretrained}): MoNuSAC, PanNuke, and Big. 
	
	\subsubsection{{Effects of model structure design for DTSeg}} Self-attention plays a crucial role in the transformer-based decoder as it aids in better aggregating of semantic features. Table \ref{ablation_sa} demonstrates the importance of the self-attention mechanism in semantic segmentation in the majority of cases, particularly when there is limited pre-training data. For instance, when the pre-training was done only on PanNuke, self-attention significantly outperformed non-self-attention at annotation levels of 1/10 (p-value\textless 0.1) and 1/5 (p-value\textless 0.05).
 
 Due to the varying receptive fields associated with different feature maps from the pre-trained diffusion model, a parallel processing approach was adopted to handle them separately. Notably, in Table \ref{ablation_feature_blocks}, the term ``Serial" refers to the concatenation of all feature blocks along the channel dimension, followed by processing with a transformer. On the other hand, the term ``Parallel" indicates the individual processing of each feature block with a transformer, followed by the concatenation of all feature blocks along the channel dimension.
Table \ref{ablation_feature_blocks} confirms that DTSeg performs more stably and achieves better results with parallel feature block processing, particularly when a smaller number of images were used during the pre-training.
	
\subsubsection{{Effects of collaborative learning with different collaborators}} As illustrated in Table \ref{ablation_collaboration}, regardless of the chosen supervised baseline, collaborative learning proved to improve the segmentation performance.  Also, collaborative learning between supervised baseline and DTSeg outperformed other collaborative learning frameworks.  Table \ref{collaboration_semi} indicates that semi-supervised training is ineffective for collaborative learning, particularly when only a small portion is labeled. The collaborative learning framework in this paper is trained using labeled cell nucleus images. However, the advantage of semi-supervised training lies in its utilization of unlabeled data rather than labeled data. Consequently, when relying solely on labeled data for collaborative learning, semi-supervised training often falls short of surpassing supervised baselines.

	\section{Conclusion}
 In this work, a large-scale unsupervised diffusion pre-training-based semi-supervised cell nuclei semantic segmentation framework has been proposed, named as DTSeg. it has been demonstrated that the unsupervised pre-training of a latent diffusion model can significantly enhance downstream semantic segmentation tasks when a large number of labeled data is not available for training.
Collaborative learning is further included to improve the performance of the proposed framework for domain-specific issues like limited data for pre-training and OOD cases. Extensive experiments and ablation studies on four publicly available cell segmentation datasets have been performed to evaluate the efficacy of our proposed method. The results have demonstrated that the diffusion model is an effective 'semi-supervised learner' for segmentation, and the strategy of large-scale pre-training can be helpful for both in-distribution and out-of-distribution test cases.

\bibliographystyle{unsrt}  
\bibliography{references}

\end{document}